\documentstyle[preprint,aps]{revtex}

\tighten
\begin{document}

\title{
      Microscopic origins of effective charges in the shell model
}
\medskip

\author{
        Petr Navr\'atil\footnote{On leave of absence from the
   Institute of Nuclear Physics,
                   Academy of Sciences of the Czech Republic,
                   250 68 \v{R}e\v{z} near Prague,
                     Czech Republic.},
     Michael Thoresen,
     and Bruce R. Barrett
        }

\medskip

\address{
                   Department of Physics,
                   University of Arizona,
                   Tucson, Arizona 85721
}

\maketitle

\bigskip

\begin{abstract}
We use a large-scale $6\hbar\Omega$ calculation for $^6$Li with
microscopically derived two-body interaction
to construct the $0\hbar\Omega$ $0p$-shell effective hamiltonian,
electric quadrupole, and magnetic dipole operators. 
While the E2 and M1 $6\hbar\Omega$ operators are one-body
operators with free nucleon charges, the effective operators
are two-body operators with substantially different
renormalization for the isoscalar and isovector matrix elements,
especially for the E2 operator. 
We show that these operators can be very well approximated
by one-body operators provided that effective proton {\it and}
neutron charges are used. The obtained effective
charges are compatible with those used
in phenomenological shell-model studies.
The two-body part of the effective operators is estimated.
\end{abstract}

\bigskip
\bigskip
\bigskip

\narrowtext



Considerable effort has been devoted to derive the
effective interaction used in the shell-model calculations
from the nucleon-nucleon interaction \cite{B67,HJKO95,EO77}. 
On the other hand, much less work has been done to understand the
effective operators employed in calculating different nuclear, 
usually electromagnetic, properties. In particular,
a microscopic derivation of effective operators has been
only partially succesful. It is well known that effective
proton and neutron charges must be employed to describe
the E2 transitions and moments. These charges are quite
different from the free nucleon charges, typically the values of
$e_{\rm eff}^{\rm p}\approx 1.5 e$ and 
$e_{\rm eff}^{\rm n}\approx 0.5 e$
are obtained for both light and heavy nuclei.
Attempts to derive these charges microscopically, usually by
perturbation \cite{EO77}, or by an ``expanded shell-model''
approach \cite{FP78}, yielded much smaller values.
It should be noted that these effective charges correspond
to a severely truncated single-major-shell, 
or $0\hbar\Omega$, space.
The question arises as to what causes the nucleon properties
to change so significantly, is it mostly
the result of the space truncation or the fact that nucleons
are bound? Also
the non-nucleonic degrees of freedom may play an important role.
Other interesting questions are: how important are the
higher than one-body parts of the effective operators and what
is the $j$-dependence of effective charges?

In this contribution we investigate how severe space truncation
affects the electromagnetic operators. We use a large-space
$6\hbar\Omega$ shell-model calculation for $^6$Li, 
with a microscopically derived two-body interaction, to
construct an effective hamiltonian and effective electromagnetic
operators, which will exactly reproduce the $6\hbar\Omega$ results 
in the $0p$-shell for the $(0s)^4(0p)^2$ dominated states.
This enables us to compare the resulting effective operators 
with the bare one-body $0p$-shell operators and to extract
the relevant effective charges, which allow us to determine 
the amount of
renormalization, to study their $j$-dependence, and, eventually,
to quantify the two-body content of the effective operators.
Also we perform the same derivation from the corresponding 
$4\hbar\Omega$ calculation to study the dependence on the space size
and compare the rate of convergence for the effective hamiltonian
and the effective operators. 

Recently, large-basis no-core shell-model calculations have been 
performed \cite{ZVB,ZBVHS}. 
In these calculations all nucleons are active,
which simplifies the effective interaction, as no hole states
are present. In the approach taken, the effective interaction is
determined microscopically from the nucleon-nucleon interaction
for a system of two nucleons and subsequently
used in the many-particle calculations. To take into account a
part of the many-body effects, a multi-valued effective interaction 
approach was introduced \cite{ZBVHS}, which uses different sets
of the effective interaction for different $\hbar\Omega$ excitations.
In the latest application
of the no-core approach, we derived starting-energy-independent
hermitian two-body effective interactions from the Reid 93 
nucleon-nucleon potential \cite{SKTS} and applied them
in the multi-valued approach to $A=3-6$ nuclei \cite{NB96prc}.
In this study we use the
results of this calculation for $^6$Li presented in 
the third column of
Table IV of Ref. \cite{NB96prc}. 
The many-particle calculation was done using
the Many-Fermion-Dynamics Shell-Model Code \cite{VZ94}
in the m-scheme with dimensions approaching $2\times 10^5$. 
As in the previous large-scale no-core 
shell-model calculations \cite{ZVB,ZBVHS}, a reasonable description
of the electromagnetic properties has been achieved using
free nucleon charges. 
Our aim here is to study the renormalization of
these operators, when the model space is severely truncated.

For the $0\hbar\Omega$ dominated states of $^6$Li
shown in Table IV of Ref. \cite{NB96prc}
it is possible to formulate 
an equivalent description purely in the $0p$-shell. We may
divide the basis states of the $6\hbar\Omega$ calculation into two
subspaces, using the projectors
$P$ and $Q$, $P+Q=1$. Here the $P$-space is spanned by the
states $|(0s)^4(0p)^2\rangle$. There are 10 such states in the
$M_J=0$ m-scheme calculation and 8 in the $M_J=1$ calculation, 
respectively.
The $Q$-space is then formed by the rest of the almost 200,000 states.
The effective $P$-space hamiltonian may be constructed
by employing the Lee-Suzuki starting-energy independent
similarity transformation
method \cite{LS80}, which gives the effective hamiltonian 
$PH_{\rm eff}P = PHP + PHQ\omega P$,
with the transformation operator $\omega$ 
satisfying 
$\omega=Q\omega P$. In the case when the
full space eigenvectors are known, like in our situation,
this operator may be obtained directly.
Let us denote the P-space states as $|\alpha_P\rangle$,
and those which belong to the Q-space, as $|\alpha_Q\rangle$.
Then the Q-space components of the eigenvector $|k\rangle$ of
the full-space hamiltonian can be expressed as a combination
of the P-space components with the help of the operator $\omega$
\begin{equation}\label{eigomega}  
\langle\alpha_Q|k\rangle=\sum_{\alpha_P}
\langle\alpha_Q|\omega|\alpha_P\rangle \langle\alpha_P|k\rangle \; .
\end{equation}
If the dimension of the model space is $d_P$, we may choose a set
${\cal K}$ of $d_P$ eigenevectors $|k\rangle$, 
for which the relation (\ref{eigomega}) 
will be satisfied. In our case we choose those states,
which have the dominant $0\hbar\Omega$ component.
Under the condition that the $d_P\times d_P$ 
matrix $\langle\alpha_P|k\rangle$ for $|k\rangle\in{\cal K}$
is invertible, which is satisfied in the present application, 
the operator $\omega$ can be determined from (\ref{eigomega}). 
Consequently, the effective hamiltonian can be constructed as follows
\begin{eqnarray}\label{effomega}
\langle \gamma_P|H_{\rm eff}|\alpha_P\rangle &=&\sum_{k\in{\cal K}}
\left[
\langle\gamma_P|k\rangle E_k\langle k|\alpha_P\rangle \right.
\nonumber \\
&& \left.
+\sum_{\alpha_Q}\langle\gamma_P|k\rangle E_k\langle k|\alpha_Q\rangle
\langle\alpha_Q |\omega|\alpha_P\rangle\right] \; .
\end{eqnarray}
It should be noted that 
$P|k\rangle=\sum_{\alpha_P}|\alpha_P\rangle\langle\alpha_P|k\rangle$
for $|k\rangle\in{\cal K}$ is a right eigenvector of (\ref{effomega})
with the eigenvalue $E_k$.
The hamiltonian (\ref{effomega}) is, in general, non-hermitian, or
more accurately quasi-hermitian. It can be hermitized by a similarity
transformation, which 
is determined from the metric operator $P(1+\omega^\dagger\omega)P$. 
The hermitian hamiltonian is then given by \cite{S82SO83}
\begin{equation}\label{hermeffomega}
\bar{H}_{\rm eff}
=\left[P(1+\omega^\dagger\omega)P\right]^{\frac{1}{2}}
H_{\rm eff}\left[P(1+\omega^\dagger\omega)P\right]^{-\frac{1}{2}} \; .
\end{equation}

Similarly, a corresponding effective operator $\hat{O}_{\rm eff}$
can be constructed for any full space, e.g., electromagnetic, operator 
$\hat{O}$ so that it exactly reproduces the full space 
results for the $P$-space eigenstates. A double-similarity 
transformation \cite{NG93} leads to the $P$-space operator
associated with the hamiltonian (\ref{effomega}) in the form
$\hat{O}_{\rm eff} = \left[P(1+\omega^\dagger\omega)P\right]^{-1}
(P+P\omega^\dagger Q)\hat{O}(P+Q\omega P)$.
The operator associated with the hermitian $P$-space hamiltonian
(\ref{hermeffomega}) is then obtained as \cite{NG93,KEHZSOK} 
\begin{eqnarray}\label{hermeffop}
\bar{O}_{\rm eff} &=& 
\left[P(1+\omega^\dagger\omega)P\right]^{-\frac{1}{2}}
(P+P\omega^\dagger Q)\hat{O}(P+Q\omega P) \nonumber \\ 
&&\times \left[P(1+\omega^\dagger\omega)P\right]^{-\frac{1}{2}}
\; .
\end{eqnarray}

Using the Eqs. (\ref{eigomega},\ref{effomega},\ref{hermeffomega}),
we constructed the effective hamiltonian, whose matrix elements,
after performing the transformation from m-scheme to $JT$ basis,
are presented in Table \ref{tab2}. In the same Table the well-known
Cohen-Kurath matrix elements \cite{CK65} are shown. 
These were obtained by 
a least-square fit to experimental binding energies, 
relative to $^4$He, and excitation energies of $A=6-16$ nuclei.
To make a meaningful comparison, the calculated  
binding energy of $^4$He, 27.408 MeV, obtained
by using the same no-core approach in the $8\hbar\Omega$ space 
\protect\cite{NB96prc},
was added to the diagonal matrix elements of our hamiltonian.
Note that by diagonalizing the hamiltonian in Table \ref{tab2}, we get
the same excitation energies as those from the $6\hbar\Omega$
calculation given in the third column of
Table IV of Ref. \cite{NB96prc}.
We also present the effective hamiltonian obtained in the
same way from a $4\hbar\Omega$ calculation. Note that the dimension
of the $4\hbar\Omega$ calculation is more than an order 
of magnitude smaller
than that of the $6\hbar\Omega$ calculation. The effective interaction
used in this calculation was obtained from the $6\hbar\Omega$
multi-valued interaction by leaving out the set corresponding
to the $6\hbar\Omega$ space. Let us mention that the change in the 
low-lying eigenenergies in the two calculations is not substantial 
and the ordering of levels is identical.
We observe that our
calculated matrix elements differ in some cases
from the phenomenological ones. Let us point out, however,
that our matrix elements provide a better description of $^6$Li
states than those of Cohen-Kurath. This can be understood as
the latter matrix elements were fitted to a large number of nuclei
across the entire $0p$-shell.

Our primary aim is to derive $0p$-shell effective 
electromagnetic operators. In the full-space calculation 
we employed the one-body E2 and M1 operators
\begin{mathletters}\label{E2M1}
\begin{eqnarray}
T^{({\rm E2})}&=& e^{\rm p} \sum_{i=1}^A (\textstyle{\frac{1}{2}}+t_{zi})
r^2_i Y^{(2)}(\Theta_i) \nonumber \\
&&+ e^{\rm n} \sum_{i=1}^A (\textstyle{\frac{1}{2}}-t_{zi})
r^2_i Y^{(2)}(\Theta_i) \; ,  \label{E2} \\ 
T^{({\rm M1})}&=& \sqrt{\textstyle{\frac{3}{4\pi}}}\mu_{\rm N}
\sum_{i=1}^A \left[(\textstyle{\frac{1}{2}}+t_{zi})
(g^{\rm p}_l {\bf l}_i +g^{\rm p}_s {\bf s}_i) \right. \nonumber \\
&&\left. + (\textstyle{\frac{1}{2}}-t_{zi})
(g^{\rm n}_l {\bf l}_i +g^{\rm n}_s {\bf s}_i)\right] \; ,\label{M1}
\end{eqnarray}
\end{mathletters}
with the free nucleon charges $e^{\rm p}=e$, $e^{\rm n}=0$ and
free nucleon $g$-factors $g^{\rm p}_l=1$, $g^{\rm n}_l=0$,
$g^{\rm p}_s=5.586$, and $g^{\rm n}_s=-3.826$. 
The $P$-space operators are constructed by the application 
of Eq. (\ref{hermeffop}). 
We calculate the $P$-space $T^{({\rm E2})}$ operator
and separately the orbital and spin parts of the M1 operator.
The calculation is performed in the m-scheme and subsequently 
transformed to the $J,T$ basis. To get all the
reduced matrix elements, full-space calculations with $M_J=0$
and $M_J=1$ must be done.
In Table \ref{tab3} we present selected matrix elements
of pieces of (\ref{E2M1}), namely the operators
\begin{mathletters}\label{E2pMlMs}
\begin{eqnarray}  
{\rm E2}&\equiv& \sqrt{\textstyle{\frac{16\pi}{5}}}
\sum_{i=1}^A (\textstyle{\frac{1}{2}}+t_{zi})
r^2_i Y^{(2)}(\Theta_i) \; ,  \label{E2p} \\
{\rm Ml}&\equiv& \sum_{i=1}^A (\textstyle{\frac{1}{2}}+t_{zi})
{\bf l}_i \; ,   \label{Ml} \\
{\rm Ms}&\equiv& \sum_{i=1}^A (\textstyle{\frac{1}{2}}+t_{zi})
{\bf s}_i \; .  \label{Ms}
\end{eqnarray}
\end{mathletters}
These matrix elements are reduced in $J$ and for $T_z=0$.
The second column shows the matrix elements of the effective 
operators, as obtained from Eq.(\ref{hermeffop}) and the
procedure outlined above. Note that these operators,
when used with the eigenvectors of the effective hamiltonian
obtained from (\ref{hermeffomega}), whose matrix elements are 
shown in Table \ref{tab2}, give the same mean values and 
transition rates as the original one-body operators (\ref{E2pMlMs}),
when used with the $0\hbar\Omega$ dominated eigenvectors 
of the $6\hbar\Omega$ calculation.
Also note that the
effective operators are two-body operators unlike the full-space  
original operators.

Let us first discuss the E2 operator. In the third column
of Table \ref{tab3} the reduced matrix elements of the operators
(\ref{E2pMlMs}), evaluated in the $P$-space, are shown for comparison.  
We observe, that there is a striking difference in the renormalization
of the isoscalar and isovector matrix elements of ${\rm (E2)_{eff}}$. 
The former are much larger in magnitude
than the latter in comparison with the unrenormalized
values of the operator (\ref{E2p}). Apparently,
there is no chance to approximate the effective operator 
as (\ref{E2p}) multiplied by some effective charge. 
Instead, it is possible
to mimic the mentioned isoscalar-isovector effect by approximating
the effective operator as a combination of one-body proton 
{\it and} neutron operators with different effective charges, e.g.,
\begin{eqnarray}\label{E2peff}  
({\rm E2})_{\rm eff} &\approx& e_{\rm eff}^{\rm p} 
\sqrt{\textstyle{\frac{16\pi}{5}}}
\sum_{i=1}^A (\textstyle{\frac{1}{2}}+t_{zi})
r^2_i Y^{(2)}(\Theta_i) \nonumber \\
&&+e_{\rm eff}^{\rm n} 
\sqrt{\textstyle{\frac{16\pi}{5}}}
\sum_{i=1}^A (\textstyle{\frac{1}{2}}-t_{zi})
r^2_i Y^{(2)}(\Theta_i) \;,
\end{eqnarray}
where only valence nucleons contribute in the sums.
A better approximation may be obtained when the effective
charges become $j$-dependent, e.g., 
$\sum_{ij}e_{{\rm eff}ij}\langle i|\hat{O}|j\rangle a^\dagger_i a_j$
in the second-quantization form. We calculated the effective
charges from the reduced matrix elements
of the appropriate operators by a least-square fit. 
The resulting $j$-dependent,
as well as $j$-independent, effective charges are presented
in Table \ref{tab4}, and the corresponding reduced matrix elements
are shown in the fourth and fifth columns of Table \ref{tab3}.
First, we observe that this kind of approximation works very well.
Moreover, the pure two-body matrix elements, which cannot be
reproduced by an approximation of the type (\ref{E2peff}), are 
almost a factor of ten smaller than the largest one-body 
matrix elements.
Second, the calculated effective charges 
$e^{\rm p}_{\rm eff}=1.527 e$, $e^{\rm n}_{\rm eff}=0.364 e$
are close to the phenomenological ones mentioned in the
introduction. Third, the $j$-dependence of the effective
charges is rather moderate.

From the phenomenological studies it is well known
that the magnetic dipole transitions and moments can
be in most cases described, at least in light nuclei,
by using the operator (\ref{M1}) with small modification
of the $g$-factors. 
Also the effective orbital and spin operators
obtained in our study are much less renormalized, when
compared to the starting one-body operators, than the
electric quadrupole operator. It is reflected in the
second and third parts of Table \ref{tab3}. The 
isoscalar-isovector effect is much smaller, and, in the case
of the spin operator it is almost non-existent. Unlike the
case of the quadrupole operator, the effective dipole operator
matrix elements are, on average, smaller in comparison with
one-body ones. A perhaps surprising result is, however, 
the observation
that the orbital part is more renormalized than the spin part
and, moreover, a neutron orbital part is generated with
an effective $g$-factor $g_{l{\rm eff}}^{\rm n}=0.085$.
The proton $g$-factor is about 10\% quenched 
$g_{l{\rm eff}}^{\rm p}=0.907$. The spin part is quenched
by about 6\%, e.g., 
$g_{s{\rm eff}}^{\rm p}=0.937 g_{s}^{\rm p}+0.001 g_{s}^{\rm n}$,
and from the isospin symmetry
$g_{s{\rm eff}}^{\rm n}=0.937 g_{s}^{\rm n}+0.001 g_{s}^{\rm p}$.
As in the quadrupole operator case, the $j$-dependence
is also moderate for the magnetic dipole operator effective charges,
but the pure two-body matrix elements are relatively smaller.

The $j$-independent effective charges extracted in the same way
from the $4\hbar\Omega$ calculation are also presented in Table
\ref{tab4}. We observe that the renormalization is smaller
in this case. This reflects the fact that, for example, 
the E2 transition
rates obtained in the $4\hbar\Omega$ calculation are weaker
than those calculated in the $6\hbar\Omega$ space. Our observation 
here is that the effective hamiltonian converges more rapidly
than the E2 operator with respect to the full-space size change.

To quantify the two-body content of the
effective $0p$-shell electromagnetic operators, we evaluate the
quantity 
$R\equiv\sqrt{\sum_{i\le j}(\hat{O}_{{\rm eff},ij}
-\hat{O}_{{\rm ob- eff},ij})^2/
\sum_{i\le j}(\hat{O}_{{\rm eff},ij})^2}$, 
where the $\hat{O}_{ij}$ denote the matrix elements 
between the $0p$-shell two-body states $i$ and $j$.
In this way we estimate the part of the effective operators, which
cannot be expressed as a combination of one-body operators.
When using one-body operators with $j$-dependent charges,
we obtain a two-body contribution of 
10.1\% for E2, 3.8\% for Ml, and 3.2\% for Ms, respectively. 
For the one-body operator with $j$-independent effective charges
the two-body contributions are 
12.3\% for E2, 5.3\% for Ml, and 4.4\% for Ms, respectively. 
Clearly, the magnetic dipole operators 
are better approximated
by combinations of one-body operators.

In conclusion, 
we have shown that model-space truncation is sufficient
to generate operator renormalization, which is characterized by
effective charges compatible with those used in the phenomenological
applications. We have found that the isoscalar and isovector
parts of the operators are renormalized differently,
particularly in the case of the electric quadrupole operator.
This difference in renormalization is the source
of a non-zero neutron effective charge. 
These findings are based on
a no-core $6\hbar\Omega$ calculation for $^6$Li with
a multi-valued starting-energy-independent two-body interaction
derived microscopically from the Reid 93 nucleon-nucleon
potential, from which the $0\hbar\Omega$ 
$0p$-shell effective hamiltonian,
electric quadrupole, and magnetic dipole operators were constructed. 
The obtained effective operators are two-body operators.
We have shown, however, that they may be well approximated
by one-body operators. 
Their two-body content is about
10\% for E2 and about 4\% for M1 opertators, respectively.
We also studied the dependence of the renormalization on the
size of the full space. We observed a non-negligible difference
between the effective E2 charges extracted from the $6\hbar\Omega$
calculation in comparison to those obtained from the $4\hbar\Omega$
calculation. On the other hand, the changes in the effective
hamiltonians obtained in the two calculations are less pronounced.   

\acknowledgements{
This work was supported by the NSF grant No. PHY93-21668.
P.N. also acknowledges partial support from 
the Czech Republic grant GA ASCR A1048504.
}

\begin{table}
\begin{tabular}{cddd}
$\langle 2j_1 2j_2, J T|H|2j_3 2j_4, JT\rangle$ & Eff & Eff-4 & CK 6-16 \\
\hline
$\langle 11, 01|H| 11, 01\rangle$ &  6.772 &  7.165 &  4.88 \\
$\langle 11, 01|H| 33, 01\rangle$ & -2.756 & -3.201 & -5.32 \\
$\langle 33, 01|H| 33, 01\rangle$ &  0.493 & -0.970 &  0.52 \\
$\langle 11, 10|H| 11, 10\rangle$ &  3.999 &  4.202 &  0.28 \\
$\langle 11, 10|H| 13, 10\rangle$ & -0.776 & -0.988 & -1.39 \\
$\langle 13, 10|H| 13, 10\rangle$ & -0.788 & -1.333 & -2.64 \\
$\langle 11, 10|H| 33, 10\rangle$ &  2.086 &  2.464 &  1.09 \\
$\langle 13, 10|H| 33, 10\rangle$ & -4.107 & -4.708 & -4.02 \\
$\langle 33, 10|H| 33, 10\rangle$ &  0.815 & -0.707 &  0.12 \\
$\langle 13, 11|H| 13, 11\rangle$ &  5.780 &  5.308 &  4.76 \\
$\langle 13, 20|H| 13, 20\rangle$ &  0.199 & -1.366 & -0.32 \\
$\langle 13, 21|H| 13, 21\rangle$ &  4.303 &  3.514 &  2.76 \\
$\langle 13, 21|H| 33, 21\rangle$ &  1.377 &  1.613 &  2.21 \\
$\langle 33, 21|H| 33, 21\rangle$ &  2.694 &  1.162 &  2.61 \\
$\langle 33, 30|H| 33, 30\rangle$ & -1.842 & -3.771 & -3.42
\end{tabular}
\caption{The $0p$-shell effective hamiltonian matrix elements, in MeV,
obtained from $6\hbar\Omega$, second column, and from $4\hbar\Omega$, 
third column, calculation for $^6$Li.
The calculated binding energy of $^4$He, 27.408 MeV, obtained
by using the same method in the $8\hbar\Omega$ space 
\protect\cite{NB96prc},
was added to the diagonal matrix elements
in order to make a meaningful comparison
with the Cohen-Kurath phenomenological matrix elements
\protect\cite{CK65}, presented in the fourth column.}
\label{tab2}
\end{table}

\begin{table}
\begin{tabular}{cdddd}
$\langle \hat{O}\rangle$ & Eff & ob & ob eff-j & ob eff   \\
\hline
$\langle 11, 10|{\rm E2}| 13, 20\rangle$ &  5.916 &  2.739 &  5.539 &  5.179 \\
$\langle 11, 10|{\rm E2}| 13, 21\rangle$ & -3.101 & -2.739 & -3.256 & -3.184 \\
$\langle 13, 20|{\rm E2}| 13, 21\rangle$ &  2.137 &  2.092 &  2.321 &  2.432 \\
$\langle 13, 21|{\rm E2}| 13, 21\rangle$ & -3.612 & -2.092 & -3.605 & -3.956 \\
$\langle 13, 10|{\rm E2}| 33, 21\rangle$ & -2.481 & -1.937 & -2.303 & -2.251 \\
$\langle 13, 10|{\rm E2}| 33, 30\rangle$ & -6.295 & -3.240 & -6.554 & -6.128 \\
$\langle 33, 01|{\rm E2}| 33, 21\rangle$ & -3.942 & -2.236 & -3.854 & -4.229 \\
$\langle 33, 21|{\rm E2}| 33, 30\rangle$ &  4.122 &  3.742 &  4.152 &  4.350 \\
$\langle 11, 10|{\rm E2}| 11, 10\rangle$ &  0.501 &  0.0   &  0.0   &  0.0   \\
$\langle 11, 10|{\rm E2}| 33, 30\rangle$ &  0.649 &  0.0   &  0.0   &  0.0   \\
\hline
$\langle 11, 01|{\rm Ml}| 11, 10\rangle$ & -0.902 & -1.155 & -0.912 & -0.949 \\
$\langle 11, 10|{\rm Ml}| 11, 10\rangle$ &  1.560 &  1.633 &  1.585 &  1.621 \\
$\langle 13, 11|{\rm Ml}| 33, 10\rangle$ & -0.439 & -0.646 & -0.509 & -0.531 \\
$\langle 13, 21|{\rm Ml}| 33, 21\rangle$ & -0.603 & -0.646 & -0.586 & -0.641 \\
$\langle 33, 01|{\rm Ml}| 33, 10\rangle$ & -1.098 & -1.291 & -1.096 & -1.061 \\
$\langle 33, 30|{\rm Ml}| 33, 30\rangle$ &  3.122 &  3.055 &  3.106 &  3.032 \\
$\langle 11, 01|{\rm Ml}| 33, 10\rangle$ & -0.100 &  0.0   &  0.0   &  0.0   \\
\hline
$\langle 11, 01|{\rm Ms}| 11, 10\rangle$ &  0.289 &  0.289 &  0.280 &  0.270 \\
$\langle 11, 10|{\rm Ms}| 11, 10\rangle$ & -0.336 & -0.408 & -0.360 & -0.383 \\
$\langle 11, 10|{\rm Ms}| 13, 20\rangle$ &  0.804 &  0.913 &  0.829 &  0.856 \\
$\langle 11, 10|{\rm Ms}| 13, 21\rangle$ & -0.819 & -0.913 & -0.839 & -0.855 \\
$\langle 33, 01|{\rm Ms}| 33, 10\rangle$ & -0.618 & -0.646 & -0.620 & -0.604 \\
$\langle 33, 30|{\rm Ms}| 33, 30\rangle$ &  1.460 &  1.528 &  1.476 &  1.433 \\
$\langle 11, 01|{\rm Ms}| 33, 10\rangle$ &  0.041 &  0.0   &  0.0   &  0.0   
\end{tabular}
\caption{Selected reduced matrix elements of the proton E2,
in $\hbar/m\Omega$, and M1, in $\mu_{\rm N}$, operators.
Here, $\langle \hat{O}\rangle\equiv
\langle 2j_1 2j_2, J_1 T_1|\hat{O}|2j_3 2j_4, J_3T_3\rangle$.
In the second column the $0p$-shell effective operator matrix elements, 
obtained from the $6\hbar\Omega$ calculation for $^6$Li, are presented.
The third column shows the corresponding proton one-body (ob) operator
matrix elements. The fourth and fifth columns display the matrix 
elements of the one-body operators with $j$-dependent and
$j$-independent effective charges, respectively. These operators
are combinations of one-body proton {\it and} neutron operators. 
}
\label{tab3}
\end{table}

\begin{table}
\begin{tabular}{cdddddd}
 & $e^{\rm p}_{\frac{1}{2}\frac{3}{2}}$  
 & $e^{\rm n}_{\frac{1}{2}\frac{3}{2}}$
 & $e^{\rm p}_{\frac{3}{2}\frac{3}{2}}$
 & $e^{\rm n}_{\frac{3}{2}\frac{3}{2}}$
 & $e^{\rm p}_{\frac{1}{2}\frac{1}{2}}$
 & $e^{\rm n}_{\frac{1}{2}\frac{1}{2}}$ \\
\hline
E2 & 1.606 & 0.417 & 1.417 & 0.307 & - & - \\
Ml & 0.848 & 0.060 & 0.933 & 0.084 & 0.880 & 0.090 \\
Ms & 0.914 &-0.006 & 0.963 & 0.003 & 0.925 &-0.043 \\  
\hline
 & $e^{\rm p}_{\rm eff}$ & $e^{\rm n}_{\rm eff}$ 
 & $e^{\rm p}_{\rm eff-4}$ & $e^{\rm n}_{\rm eff-4}$ \\
\hline
E2 & 1.527 & 0.364 & 1.302 & 0.244 \\
Ml & 0.907 & 0.085 & 0.931 & 0.063 \\
Ms & 0.937 & 0.001 & 0.953 &-0.003  
\end{tabular}
\caption{Effective charges of the proton quadrupole,
magnetic orbital and magnetic spin operators, derived by least-square 
fits to the corresponding $0p$-shell effective operators obtained from
the $6\hbar\Omega$ calculation for $^6$Li. Both the $j$-dependent
and $j$-independent effective charges are shown. 
Also, the $j$-independent effective charges obtained in the same way
from the $4\hbar\Omega$ calculation are presented in the last two
columns. 
For the definition
of the effective charges see Eq. (\protect\ref{E2peff}).
}
\label{tab4}
\end{table}

\end{document}